\documentclass[10pt,nofootinbib,notitlepage,twocolumn,superscriptaddress]{revtex4-1}

\usepackage[utf8]{inputenc}
\usepackage[T1]{fontenc}
\usepackage[english]{babel}

\usepackage{times}

\usepackage{amssymb,amsmath,amsfonts,amsthm}
\usepackage{mathtools}
\usepackage{verbatim}
\usepackage{enumerate}
\usepackage{graphicx}
\usepackage{hyperref}
\usepackage{bbm}
\usepackage{booktabs}

\usepackage[caption=false]{subfig}

\usepackage{color}
\definecolor{dred}{rgb}{.8,0.2,.2}
\definecolor{ddred}{rgb}{.8,0.5,.5}
\definecolor{dblue}{rgb}{.2,0.2,.8}
\definecolor{dgreen}{rgb}{.2,0.5,.2}

\theoremstyle{plain}

\theoremstyle{definition}

\newcommand{\bra}[1]{\mbox{$\langle #1|$}}
\newcommand{\ket}[1]{\ensuremath{|#1\rangle}}

\newcommand{\be}{\begin{equation}}
\newcommand{\ee}{\end{equation}}

\def\1#1{{\bf #1}}
\def\2#1{{\cal #1}}
\def\7#1{{\mathbb #1}}

\newcommand{\bea}{\begin{eqnarray}}
\newcommand{\eea}{\end{eqnarray}}

\begin{document}

\title{Entanglement Measures in Embedding Quantum Simulators with Nuclear Spins}

\author{Tao Xin}
\affiliation{State Key Laboratory of Low-Dimensional Quantum Physics and Department of Physics, Tsinghua University, Beijing 100084, China}
\affiliation{Tsinghua National Laboratory of Information Science and Technology,  Beijing 100084, China}

\author{Julen S. Pedernales}
\affiliation{Department of Physical Chemistry, University of the Basque Country UPV/EHU, Apartado 644, 48080 Bilbao, Spain}
\affiliation{Institute for Theoretical Physics and IQST, Albert-Einstein-Allee 11, Universit\"at Ulm, D-89069 Ulm, Germany}

\author{Enrique Solano}
\affiliation{Department of Physical Chemistry, University of the Basque Country UPV/EHU, Apartado 644, 48080 Bilbao, Spain}
\affiliation{IKERBASQUE,  Basque  Foundation  for  Science,  Maria  Diaz  de  Haro  3,  48013  Bilbao,  Spain}

\author{Gui-Lu Long}
\email[Correspondence and requests for materials should be addressed to G.L.L.: ]{gllong@tsinghua.edu.cn}
\affiliation{State Key Laboratory of Low-Dimensional Quantum Physics and Department of Physics, Tsinghua University, Beijing 100084, China}
\affiliation{Tsinghua National Laboratory of Information Science and Technology,  Beijing 100084, China}
\affiliation{The Innovative Center of Quantum Matter, Beijing 100084, China}

\begin{abstract}

We implement an embedding quantum simulator (EQS) in nuclear spin systems. The experiment consists of a simulator of up to three qubits, plus a single ancillary qubit, where we are able to efficiently measure the concurrence and the three-tangle of two-qubit and a three-qubit systems as they undergo entangling dynamics. The EQS framework allows us to drastically reduce the number of measurements needed for this task, which otherwise would require full-state reconstruction of the qubit system. Our simulator is built of the nuclear spins of 4 $^{13}C$ atoms in a molecule of trans-crotonic acid manipulated with NMR techniques.  

\end{abstract}

\maketitle

{\it{Introduction .--}} Entanglement, having no classical counterpart, is one of the most distinctive features of quantum mechanics~\cite{Amico_ent,Horodecki_ent}, and it is considered to be a fundamental resource for quantum information processing and quantum communication~\cite{chuang}.  Therefore, it is not surprising, that the quantification of entanglement is a major topic for both, the theoretical and the experimental quantum information communities.  In this respect, entanglement monotones were introduced as functionals of a quantum state that take a null value for separable states, and do not increase under local operations and classical communication (LOCC)~\cite{Vidal99}. Unfortunately, it is believed that the measurement of entanglement monotones requires, in general, full-state tomography (FST) of the system of interest, something that makes it experimentally intractable in scalable quantum systems, as the number of necessary measurements for FST grows exponentially with the system size~\cite{cramer,jullien,daweixin}. For instance,  $4^n-1$ observables need to be measured to reconstruct the wave function of $n$ qubits, which pose a difficulty comparable to the classical simulation of such a wave function with ordinary computers. There have been efforts to circumvent this difficulty. A paradigmatic example is that of `entanglement witnesses' which were introduced as detectors of entanglement. In this case, the value of a physical observable indicates whether an arbitrary quantum state is entangled or not~\cite{park}, however, only witnesses for specific kinds of entanglement are known, and not universal ones. Moreover, entanglement witnesses may detect but not quantify, in general, the amount of entanglement, and therefore do not serve as a comparative tool among different entangled states. Other alternative methods rely on collective measurements on many identical copies of a given quantum state, which is experimentally demanding as well~\cite{Bovino,Huber}. On the other hand, it is known that FST is necessary for universal entanglement detection with single-copy observables~\cite{xindawei}. All in all, quantum computers and quantum simulators do not seem a priori to be efficient tools for the quantification of entanglement, at least when compared to classical computers.

Embedding quantum simulators (EQS)~\cite{Candia,Loredo} were proposed as a path to solve this conflict in the context of scalable quantum platforms. A one-to-one quantum simulator, which directly implements the wave function of interest and its dynamics in a controllable quantum system, is bound to direct detection of the entanglement present on it, if it were to measure the entanglement of the system that it simulates. However, in many situations, one is not necessarily  interested in the entanglement present in the physical system but in the entanglement predicted by the simulated model and its evolution in time. In the same manner that the numerical simulation of quantum systems is not concerned about the inexistent entanglement among the classical bits employed for the simulation. Therefore, a suitable mapping of the model of interest to the quantum simulator that exposes the entanglement of the simulated system without the necessity of FST is of interest.  EQSs provide a systematic manner to construct quantum simulators on which entanglement monotones are accessible with a reduced number of measurements when compared to FST. This is done by the addition of a single ancillary qubit and a suitable mapping of the initial state and the Hamiltonian dynamics that allows for the measurement of anti-linear operators. 

In this letter, we show the first implementation of the EQS framework with nuclear spins. We implement two EQSs: first a simulator of two qubits implemented with three nuclear spins, and second a three-qubit quantum simulator built of four nuclear spins. For the two-qubit simulator we measure the concurrence of the system as it evolves under an entangling Hamiltonian. We do this with the measurement of just two observables, as opposed to the 15 needed for FST. In the second case, we measure the three-tangle of the system, which is an entanglement monotone that detects  genuine tripartite entanglement in a three qubit system. In this case, the necessary observables are reduced from the 63 needed for full-state reconstruction to only 6 observables. 
\\

{\it{EQS and Entanglement Monotones .--}} For a given Hamiltonian of interest, we assume a matrix representation that can be decomposed in its real and imaginary parts as $H=A+iB$, where $A$ and $B$ are real matrices, which are respectively Hermitian, $A=A^\dag$, and anti-Hermitian, $B=-B^\dag$. If we consider an arbitrary initial state $\ket{\phi(0)}$, a one-to-one quantum simulator would directly implement it in a physical system and evolve it under Hamiltonian $H$. In contrast, an EQS implements it in an enlarged Hilbert space onto which both, the initial state and the Hamiltonian, are mapped. The initial state is mapped as $\ket{\Phi(0)}=\ket{0} \otimes \text{Re}\ket{\phi(0)}+\ket{1} \otimes \text{Im}\ket{\phi(0)}$ and the Hamiltonian as $H'=i\sigma_0\otimes B-\sigma_y\otimes A$, where $\sigma_0$ is a $2\times2$ identity matrix and $\sigma_{x,y,z}$ are Pauli matrices. Under this mapping, which only requires the addition of one ancillary qubit, regardless of the size of the simulated system, expectation values of anti-linear operators can be retrieved with the measurement of two observables. Anti-linear operators take the form $OK$, where $O$ is an observable and $K$ is the complex conjugation operator, which acts on the vector elements of a ket state by complex conjugating them, $K \ket{\phi}=\ket{\phi^*}$. Anti-linear operators are not Hermitian and therefore they are not observables, generally requiring full-state reconstruction of the quantum state of a system to compute their expectation value. However, under the mapping introduced above, anti-linear operators can be efficiently computed in an EQS according to the relation
 \begin{equation}
\bra{\phi(t)}O\ket{\phi^*(t)}=\bra{\Phi(t)}\sigma_z\otimes O\ket{\Phi(t)}-i\bra{\Phi(t)}\sigma_x\otimes O\ket{\Phi(t)}.
\end{equation}

It is known that entanglement monotones for qubit systems can be systematically constructed from anti-linear operators~\cite{Siewert05}. For instance, the concurrence, which is a two-qubit entanglement monotone, can be represented as $\mathcal{C}=\bra{\phi(t)}\sigma_y\otimes \sigma_y\ket{\phi^*(t)}$. In an embedding quantum simulator, this would be retrieved from the expectation values of observables $\sigma_z\sigma_y\sigma_y$ and $\sigma_x\sigma_y\sigma_y$ in the enlarged space,
 \begin{equation}
\mathcal{C}=\left| \left \langle \sigma_z\otimes \sigma_y\otimes \sigma_y \right \rangle-i\left \langle\sigma_x\otimes \sigma_y\otimes \sigma_y \right \rangle\right|,
\end{equation}
reducing the number of required observables to 2, from the 15 required to do FST of a two-qubit system.

As an another paradigmatic example, we can mention the three-qubit entanglement monotone three-tangle, which can be defined in terms of anti-linear operators as
 \begin{equation}
\mathcal{E}_3=\left| -\left \langle O_1K \right \rangle^2 +\left \langle O_2K \right \rangle^2+\left \langle O_3K \right \rangle^2\right|,
\end{equation}
with $O_1=\sigma_0\sigma_y\sigma_y$, $O_2=\sigma_x\sigma_y\sigma_y$, and $O_3=\sigma_z\sigma_y\sigma_y$. Each of the anti-linear operators can be mapped onto two Hermitian operators in the EQS, which makes three-tangle accesible with the measurement of just six observables, as opposed to the $63$ required to do FST of 3 qubits: $\sigma_z\sigma_0\sigma_y\sigma_y$, $\sigma_x\sigma_0\sigma_y\sigma_y$, $\sigma_z\sigma_x\sigma_y\sigma_y$, $\sigma_x\sigma_x\sigma_y\sigma_y$, $\sigma_z\sigma_z\sigma_y\sigma_y$, and $\sigma_x\sigma_z\sigma_y\sigma_y$. 
\\

{\it{Experimental realisation in NMR .--}} For a proof-of-principle demonstration of EQSs in spin systems, we choose a platform of verified controllability and precision like NMR~\cite{Jones}. Although the potentiality of NMR platforms to scale up to relevant system sizes is unclear, it is important to demonstrate the working principles of EQSs in spin systems. This opens the door to the implementation of EQSs in other more scalable spin-based quantum platforms that are as well manipulated with NMR or analogous techniques. This includes, NV-centers in diamond crystals~\cite{Hollenberg13}, hyperfine-qubits in ion traps~\cite{Wineland03, Pedernales14}, or color centers in 2D materials~\cite{Plenio17}. 

In our experiment, we have used four qubits in a sample of $^{13}C$-labeled trans-crotonic acid dissolved in d6-acetone. The 4-qubit quantum simulator is implemented with the nuclear spins of 4 carbon atoms of the trans-crotonic acid molecule labeled from C1 to C4, after canceling their coupling to the methyl group M, and to the hydrogen atoms labeled as H1 and H2~\cite{sma}.  All experiments were carried out on a Bruker AVANCE 400MHz spectrometer at room temperature. The Hamiltonian of our system under the weak coupling approximation can be written as
\begin{align}\label{Hamiltonian}
\mathcal{H}_{\text{int}}=\sum\limits_{j=1}^4 {\pi (\nu _j-\nu _0) } \sigma_z^j  + \sum\limits_{j < k,=1}^4 {\frac{\pi}{2}} J_{jk} \sigma_z^j \sigma_z^k,
\end{align}
where $\nu_j$ and $\emph{J}_{jk}$ are the chemical shifts and the J-coupling strengths, respectively. $\nu _0$ is the reference frequency of $^{13}$C channel in the NMR platform. The detailed structure and parameters of the Hamiltonian can be found in Ref.~\cite{sma}. 

We initialise the system in a pseudo-pure state (PPS), which is the pure state of interest $| 0 0 0 0 \rangle $ only with probability $\epsilon$ and a maximally mixed state otherwise. This is represented with the density matrix $\rho_{0000}=(1-\epsilon)\sigma_0^{\otimes 4}/16+\epsilon\ket{0000}\bra{0000}$, where the polarisation $\epsilon$ takes the value $\epsilon=10^{-5}$ in our experiment. Conveniently enough, the expectation value of any observable measured for such a state will be that corresponding to the state $\ket{0000}$ and its time evolution, as the identity part of the state does not evolve nor contribute to the NMR signals. To generate this PPS from the initial thermal state, we have used the spatial averaging technique~\cite{cory1,cory2,dawei2}, and then performed FST~\cite{Leskowitz, JSlee} to benchmark the quality of our PPS.  We found that the initialisation fidelity was of $98.77\%$, setting the ground for reliable subsequent simulations.

\begin{figure*}[htb]
\begin{center}
\includegraphics[width= 1.4\columnwidth]{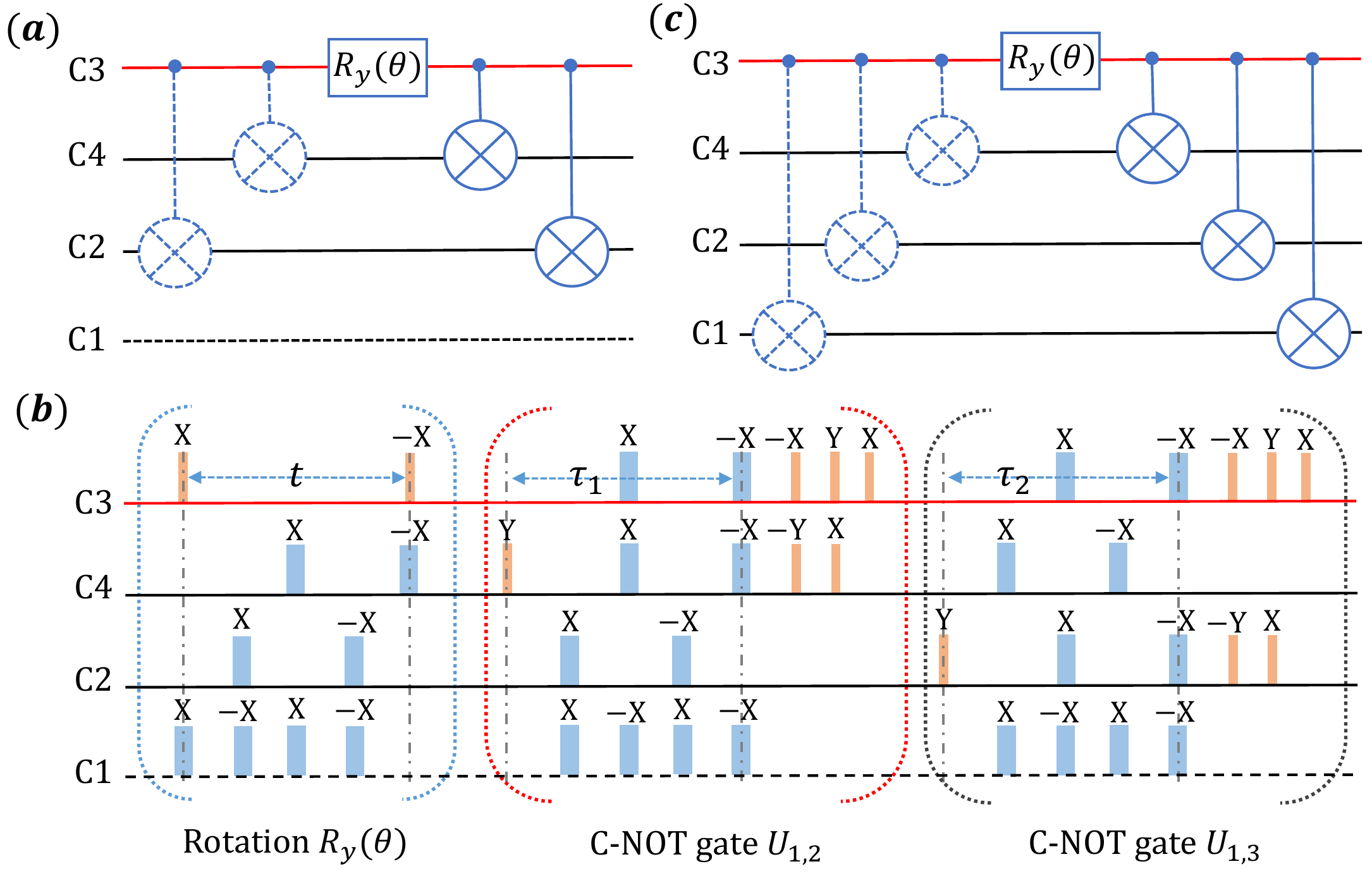}
\end{center}
\setlength{\abovecaptionskip}{-0.00cm}
\caption{\footnotesize{\textbf{Quantum circuit and corresponding NMR pulse sequence.} (a) Quantum circuit consisting of four C-NOT gates and one local rotation $R_y(\theta)$, which implements the evolution associated to Hamiltonian $H=-\omega\  \sigma_y \otimes \sigma_x \otimes \sigma_x$. The upper (red) line represents the ancillary qubit in the EQS, which is held by the nuclear spin of atom C3. Black solid and dotted lines represent the work and idle qubits, respectively. The dotted C-NOT gates can be avoided for initial states of the form $\ket{0000}$. (b) NMR pulse sequence corresponding to the circuit in (a). The orange and blue rectangles represent, respectively, $\pi/2$ and $\pi$ pulses around the directions indicated on top of them. Parameters $\tau_{1,2}$ take values $\tau_1=1/2J_{C3,C4}$ and $\tau_2=1/2J_{C3,C2}$ . (c) Quantum circuit for the implementation of Hamiltonian $H=-\omega\  \sigma_y\otimes \sigma_x\otimes \sigma_x\otimes \sigma_x$, consisting of six C-NOT gates and one local rotation $R_y(\theta)$ . }} \label{circuit}
\end{figure*}
\begin{figure}[htb!]
\begin{center}
\includegraphics[width= 0.85\columnwidth]{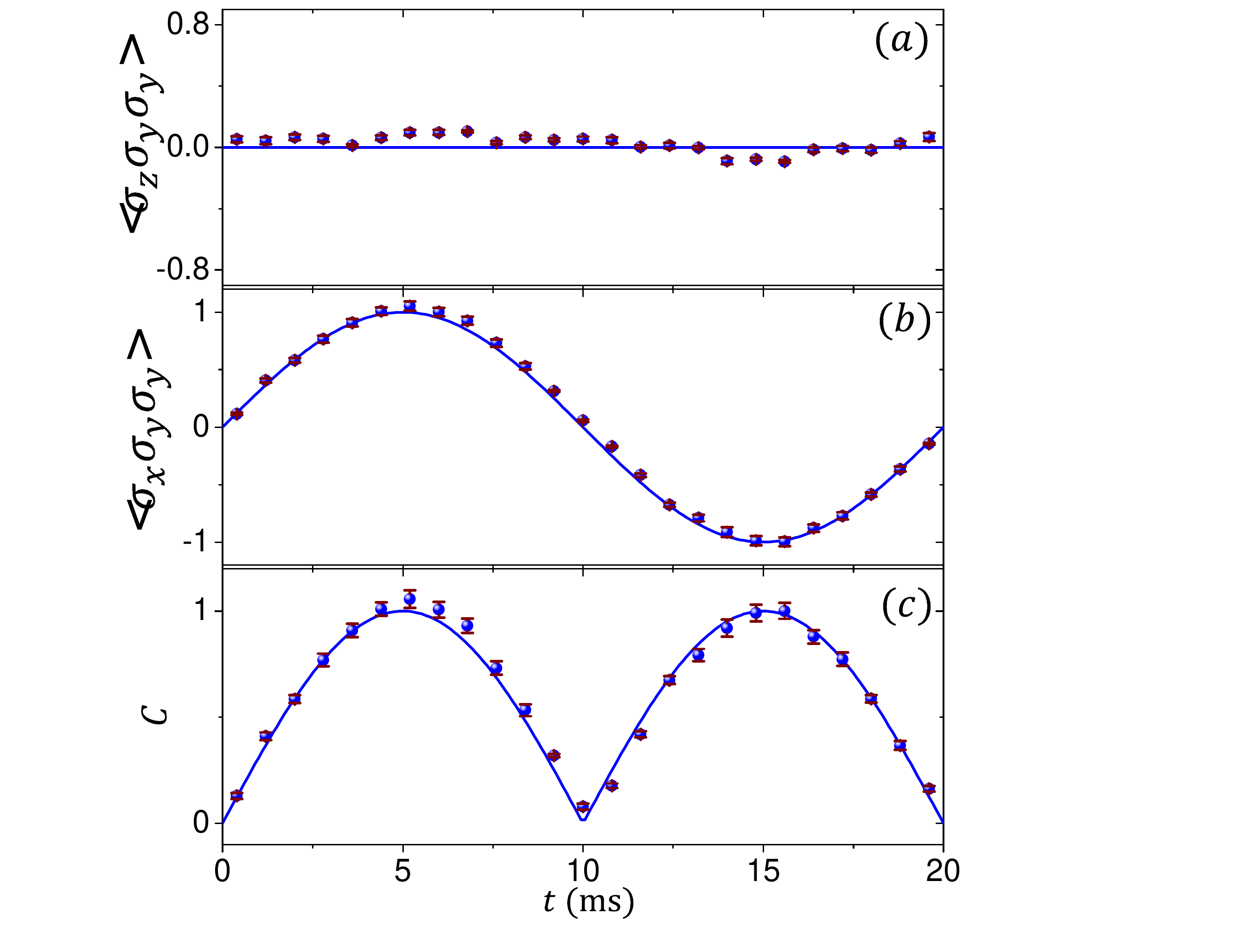}
\end{center}
\setlength{\abovecaptionskip}{-0.00cm}
\caption{\footnotesize{\textbf{Experimental results for the evolution of concurrence, $\mathcal{C}(t)$.} (a) and (b) show the time evolution of the expectation values of the EQS observables $\sigma_z\sigma_y\sigma_y$ and $\sigma_x\sigma_y\sigma_y$, respectively. (c) Reconstructed concurrence $\mathcal{C}(t)$ of the simulated model from the values of the measured <$\sigma_z\sigma_y\sigma_y$> and <$\sigma_x\sigma_y\sigma_y$>. Dots represent experimental data and lines stem from theory predictions. The error bars are calculated from the estimated imperfections of the GRAPE pulses, PPS preparation and T2-decoherence effects.}}\label{ent_1}
\end{figure}

For a first experiment, we consider a small toy model consisting of the two-qubit initial state $\ket{\phi_2(0)}=\ket{00}$ evolving under the entangling Hamiltonian ($\hbar=1$) $H_2= \omega \sigma_x\otimes \sigma_x$, with $\omega=(2\pi)\times25$Hz. The common method to track the evolution of concurrence for such a small model would involve performing FST of the evolved state $\ket{\phi_2(t)}=e^{-iH_2t}\ket{00}$ at a collection of times $t_i$, which would require the measurement of 15 observables at each time $t_i$. Then with the whole wave function concurrence would be computed according to $\mathcal{C}=\left| \bra{\phi(t)}\sigma_y \otimes \sigma_y\ket{\phi^*(t)}\right|$, which for this specific system can be shown to be $\mathcal{C}=\left| \sin{2 \omega t}\right|$.

Using the EQS formalism, the problem is recast into the initial state $\ket{\Phi_2(0)}=\ket{000}$ evolving under Hamiltonian  ${\tilde{H}_2=-\omega \sigma_y\otimes \sigma_x\otimes \sigma_x}$.  In Fig.~\ref{circuit} (a), we show the quantum circuit that implements such an evolution, which includes four controlled-NOT gates and one local rotation $R_y(\theta)=\exp(-i\theta\sigma_y/2)$ acting on the ancillary qubit with $\theta=-2\omega t $. Considering that the initial state $\ket{000}$ is unaffected by the first two controlled-NOT gates, one can reduce the circuit to that one indicated by the continuous lines in Fig. \ref{circuit} (a), disregarding the diagram parts represented with discontinuous lines. Controlled-NOT gates $U_{a,b}$, with qubit $a$ and $b$ representing the control and target qubits, respectively, can be decomposed into a suitable form for their implementation in NMR, consisting of local rotations and J-coupling kind evolutions~\cite{xin15},
\begin{eqnarray}
U_{a,b} = \sqrt{i}R^a_z(\frac{\pi}{2})R^b_z(-\frac{\pi}{2})R^b_x(\frac{\pi}{2})U(\frac{1}{2J})R^b_y(\frac{\pi}{2}).
\label{decompose}
\end{eqnarray}
Here, $U(\frac{1}{2J})$ is the $J$-coupling evolution $e^{-i\pi\sigma^a_z\sigma^b_z/4}$. Moreover, any $z$-rotation $R_z(\theta)$ can be decomposed in terms of rotations around the $x$ and $y$ axes, $R_z(\theta)=R_y(\pi/2)R_x(-\theta)R_y(-\pi/2)$.  Local rotations $R_y(\theta)$ can be realised by setting the reference frequency $\nu_0$ to satisfy the condition $\nu_3-\nu_0=-50$Hz, and using refocusing pulses to cancel the phase accumulated on the unaddressed $^{13}$C because of the offset~\cite{refc}. The specific pulse sequence consisting exclusively of local rotations and J-coupling evolutions is illustrated in Fig. \ref{circuit}(b). Because selective excitations are usually imperfect in homonuclear systems and the effect of too many pulses is accumulative resulting in a snowball effect of imprecisions, we choose to pack up all the pulses together and implement the simulation via the GRadient Ascent Pulse Engineering (GRAPE) technique~\cite{Khaneja, Ryan}. The GRAPE approach provides a 15ms shaped-pulse width and over $99.5\%$ fidelity for the whole package of pulses.

\begin{figure*}[htb]
\begin{center}
\includegraphics[width= 1.6\columnwidth]{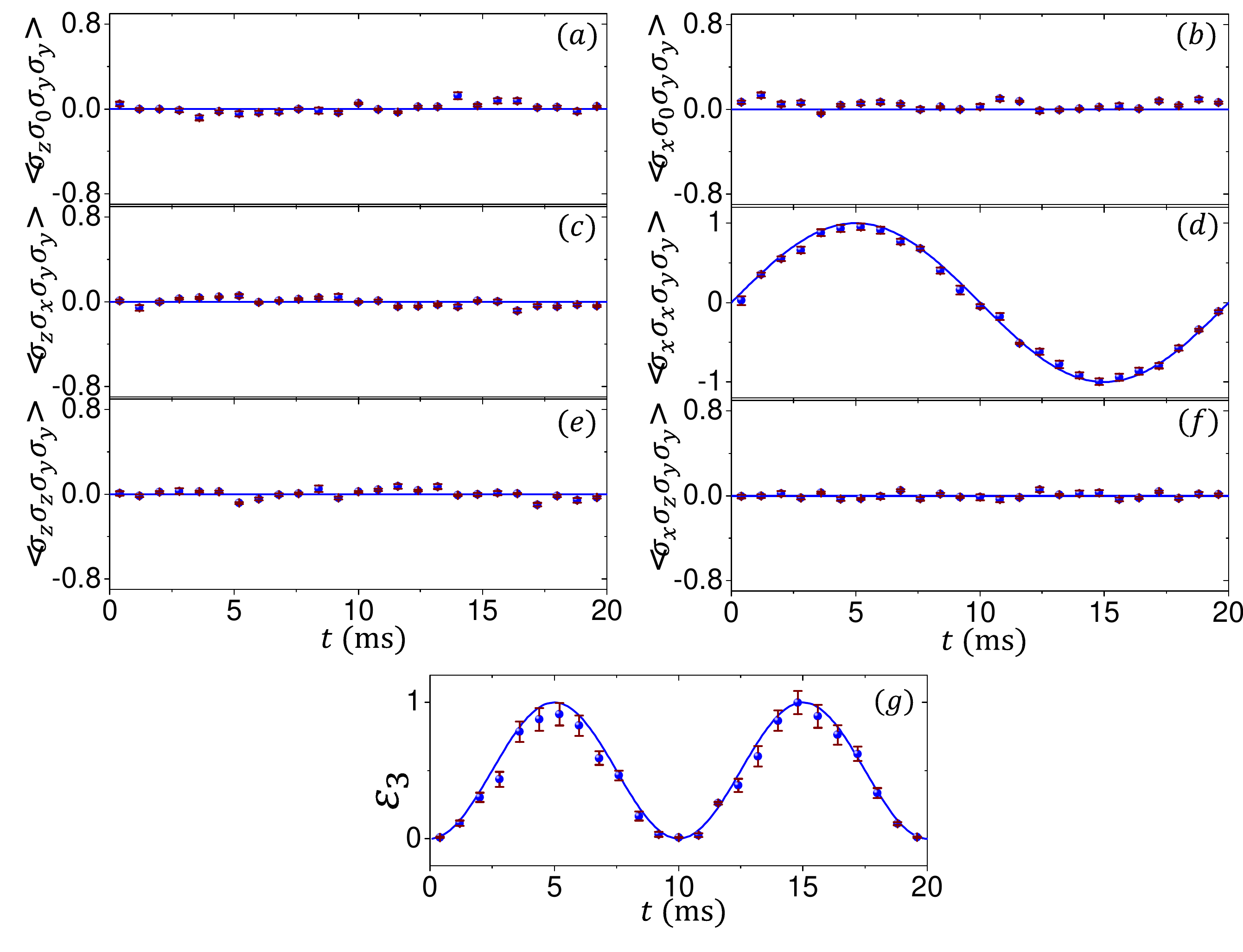}
\end{center}
\setlength{\abovecaptionskip}{-0.00cm}
\caption{\footnotesize{\textbf{Experimental results for the evolution of three-tangle, $\mathcal{E}_3$.} (a)-(f) show the expectation values of observables $\sigma_z\sigma_0\sigma_y\sigma_y$, $\sigma_x\sigma_0\sigma_y\sigma_y$, $\sigma_z\sigma_x\sigma_y\sigma_y$, $\sigma_x\sigma_x\sigma_y\sigma_y$, $\sigma_z\sigma_z\sigma_y\sigma_y$, and $\sigma_x\sigma_z\sigma_y\sigma_y$, respectively.  Plot (g) provides the result of the time evolution of three-tangle $\mathcal{E}_3(t)$ computed from the measurements of the previous six observables. Dots are experimental points and lines theory predictions.The error bars are estimated from the noise introduced by the GRAPE pulses, imperfect PPS preparation and T2-decoherence effect.} }\label{ent_2}
\end{figure*}

In our setup, we can measure expectation values of observables of the forms $\sigma_{x,y} \otimes \sigma_{0,z} ^{\otimes n-1} $ and $\sigma_{x,y} \otimes \ket{k}\bra{k} ^{\otimes n-1} $ by measuring the Free Induction Decay (FID) signal at the end of the protocol. Here, $n$ is the total number of system qubits and $k$ can take values $0$ and $1$. For the simulation of this first model, we only need three physical qubits. Therefore, we leave qubit C4 as an spectator qubit that does not take part in the dynamics, as indicated in Fig. \ref{circuit}, and we handle the data in the subspace associated to the initial state $\ket{0}$ of qubit C4. This means that the two observables of interest  $\sigma_z\sigma_y\sigma_y$ and $\sigma_x\sigma_y\sigma_y$ are retrieved from the expectation values of the four-qubit operators $\sigma_z\sigma_y\sigma_y \otimes \ket{0}\bra{0}$ and $\sigma_x\sigma_y\sigma_y\otimes \ket{0}\bra{0}$. In order to measure these operators, we perform rotations $\{YXXI,IXXI\}$ before the measurement of the FID signal, which results in transformations

\begin{equation}
\begin{array}{l}
\sigma_z\sigma_y\sigma_y \otimes \ket{0}\bra{0}~ \stackrel{\large YXXI}{\longrightarrow}~\sigma_x\sigma_z\sigma_z \otimes \ket{0}\bra{0},\\
\sigma_x\sigma_y\sigma_y \otimes \ket{0}\bra{0}~  \stackrel{\large IXXI}{\longrightarrow}~\sigma_x\sigma_z\sigma_z \otimes \ket{0}\bra{0},
\end{array}
\label{E0E1}
\end{equation}
where $X=\exp(-i\sigma_x\pi/4)$ and $Y=\exp(-i\sigma_y\pi/4)$. In this manner, the expectation values of interest are directly obtained from the experimental spectrum. The pulses corresponding to these last rotations in the measurement process are again realised using the GRAPE technique, which in this case results in a 1ms shaped-pulse of fidelity $99.5\%$. We consider $25$ temporal points ranging from $0.4$ms to $19.6$ms with an increment of $0.8$ms. For each time $t_i$, we carry out the evolution of the embedding quantum simulator  twice, and we measure after each of the evolutions the expectation value of one of the observables <$\sigma_z\sigma_y\sigma_y$> and <$\sigma_x\sigma_y\sigma_y$>. The results and their comparison to theoretical predictions are shown in Fig. \ref{ent_1}.

Errors contained in the final values of the concurrence have contributions from different origins. On the one hand, we have the imperfect initialisation of the PPS, which is estimated in an initial state infidelity of $1.30\%$. On the other hand, we have contributions from imprecisions and inhomogeneities of the GRAPE pulses. Also, magnetic field fluctuations leading to decoherence processes of the qubit systems, are expected to contribute to the final deviations of the results from their ideal values. In this respect, error bars in Fig.~(\ref{ent_2}) where computed from the comparison of the experimental expectation values with numerically simulated ones, where the noise contribution of all the mentioned sources was taken into account~\cite{sma}.

We consider a somewhat more involved case now with the simulation of a three-qubit entangling dynamics, which consists of the initial state $\ket{\phi_3(0)}=\ket{000}$ evolving under Hamiltonian ${H_3= \omega\  \sigma_x\otimes \sigma_x\otimes \sigma_x}$, with $\omega=(2\pi) \times 25$Hz. Following the same recipe introduced in the analysis of the previous case, the EQS for such a model consists of Hamiltonian $\tilde{H}_3=-\omega \ \sigma_y\otimes \sigma_x\otimes \sigma_x\otimes \sigma_x$ acting on the initial state $\ket{\Phi_4(0)}=\ket{0000}$.  Figure~\ref{circuit} (c) illustrates the corresponding quantum circuit, which includes six controlled-NOT gates and a local operation $R_y(\theta)=\exp(-i\theta\sigma_y/2)$, with $\theta=- 2 \omega t $, acting on the ancillary qubit. Based on the same considerations as those of the previous experiment, the first three controlled-NOT gates can be disregarded and the remaining pulse sequence packed up and implemented with the GRAPE technique, which for this case results in a 30ms shaped-pulse with over $99.5\%$ fidelity. Then, we measure the expectation value of the $6$ observables of interest: $\sigma_z\sigma_0\sigma_y\sigma_y$, $\sigma_x\sigma_0\sigma_y\sigma_y$, $\sigma_z\sigma_x\sigma_y\sigma_y$, $\sigma_x\sigma_x\sigma_y\sigma_y$, $\sigma_z\sigma_z\sigma_y\sigma_y$, and $\sigma_x\sigma_z\sigma_y\sigma_y$. For these, and in a similar fashion to the procedure followed in the previous experiment, we transform our final state under the rotations $\{IIXX, YIXX, Y\bar{Y}XX,I\bar{Y}XX\}$ in order to map the expectation values of interest to the measured FID signal at the end of the protocol,
\begin{align}
\sigma_x\sigma_0\sigma_y\sigma_y, \sigma_x\sigma_z\sigma_y\sigma_y~& \stackrel{\large IIXX}{\longrightarrow}~\sigma_x\sigma_0\sigma_z\sigma_z, \sigma_x\sigma_z\sigma_z\sigma_z,\\\nonumber
\sigma_z\sigma_0\sigma_y\sigma_y, \sigma_z\sigma_z\sigma_y\sigma_y~&  \stackrel{\large YIXX}{\longrightarrow}~\sigma_x\sigma_0\sigma_z\sigma_z, \sigma_x\sigma_z\sigma_z\sigma_z,\\\nonumber
\sigma_z\sigma_x\sigma_y\sigma_y~& \stackrel{\large Y\bar{Y}XX}{\longrightarrow}~ \sigma_x\sigma_z\sigma_z\sigma_z, \\\nonumber
\sigma_x\sigma_x\sigma_y\sigma_y~& \stackrel{\large I\bar{Y}XX}{\longrightarrow}~ \sigma_x\sigma_z\sigma_z\sigma_z.\\\nonumber
\end{align}
Here, $\bar{Y}=\exp(i\sigma_y\pi/4)$ and $I$ is the identity operation. The GRAPE technique is as well used to implement this last sequence of pulses by applying a 1ms shaped-pulse with a fidelity of $99.5\%$. We consider the same temporal points as those of the previous experiment. In Fig.~\ref{ent_2}, we presents the results of these experiments and the corresponding three-tangle $\mathcal{E}_3$ computed with them. 

{\it{Conclusion .--}} If scalable quantum simulators and quantum computers are to be used as a tool in the analysis of entanglement and its dynamics, they will unavoidably need to be designed under suitable mappings that guarantee that entanglement measures can be efficiently retrieved. The EQS paradigm offers a mapping which drastically reduces the number of observables that codify this information, with a minimum added complexity in the initialisation and dynamics of the quantum simulator. Here, we validate these ideas with two experiments in nuclear spins controlled with NMR techniques. Our experimental results show a high degree of correspondence with the theory predictions, opening the door to the experimental field of EQS in spin-based platforms.

{\bf Acknowledgments .--} T. X. and G.-L. L. are grateful to the following funding sources: National Natural Science Foundation of China under Grants No. 11175094 and No. 91221205; National Basic Research Program of China under Grant No. 2015CB921002. J. S. P. and E. S. acknowledge financial support from grants: Spanish MINECO/FEDER FIS2015-69983-P and Basque Government IT986-16.

\clearpage

\onecolumngrid

\section*{Supplemental Material for \\ ``Entanglement Measures in Embedding Quantum Simulators with Nuclear Spins''}

Further experimental details, as well as more insights on the employed techniques, are provided in this Supplemental Material. 
\\

\begin{figure}[htb]
\begin{center}
\includegraphics[width= 0.6\columnwidth]{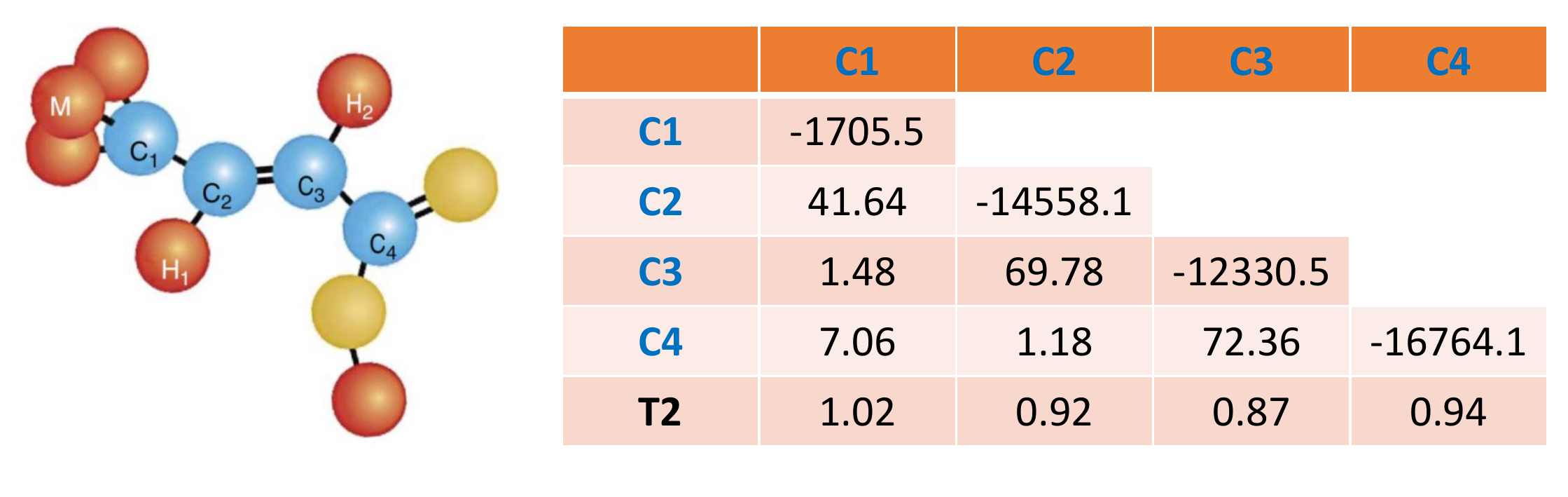}
\end{center}
\setlength{\abovecaptionskip}{-0.00cm}
\caption{\footnotesize{\textbf{Molecular structure and Hamiltonian parameters of $^{13}$C-labeled trans-crotonic acid.} In experiments, C$_1$, C$_2$, C$_3$ and C$_4$ are used as a four-qubit simulator. In the table, the chemical shifts and J-couplings (in Hz) are presented by the diagonal and off-diagonal elements, respectively. The last row of the table shows T$_{2}$ (in seconds).}}\label{molecule}
\end{figure}

{\it{\bfseries{Experimental samples}}}--In the experiments we have employed a sample of $^{13}$C-labeled trans-crotonic acid dissolved in d6-acetone, as indicated in the main text. In figure~\ref{molecule} we give a pictorial representation of the molecule structure together with the values of some relevant parameters. 
\\

{\it{\bfseries{PPS preparation}}}--Considering that our sample is composed of four $^{13}$C atoms, which makes it a homonuclear system, we can regard the gyromagnetic ratios of all the nuclear spins the same and describe the initial thermal equilibrium state as
\begin{align}\label{thermal}
\rho_{thermal}=\frac{\sigma_0^{\otimes 4}}{2^4}+\epsilon\sum _{i=1}^4 \sigma_z^i,
\end{align}
where $\epsilon\approx 10^{-5}$ represents the polarisation at room temperature. The spatial averaging technique is used to initialised our simulator in the PPS 
\begin{align}\label{pps}
\rho_{0000}=\frac{1-\epsilon}{16}\sigma_0^{\otimes 4}+\epsilon\ket{0000}\bra{0000}.
\end{align}
Figure~\ref{pps} shows the reconstructed real and imaginary parts of the PPS density matrix, where only the deviation of the state from the maximally mixed part is detectable. From these measurements a fidelity of $98.77\%$ is computed between the target pure state $\ket{0000}\bra{0000}$ and the experimentally sensitive part of the constructed PPS.
 \begin{figure*}[htb]
\begin{center}
\includegraphics[width= 0.8\columnwidth]{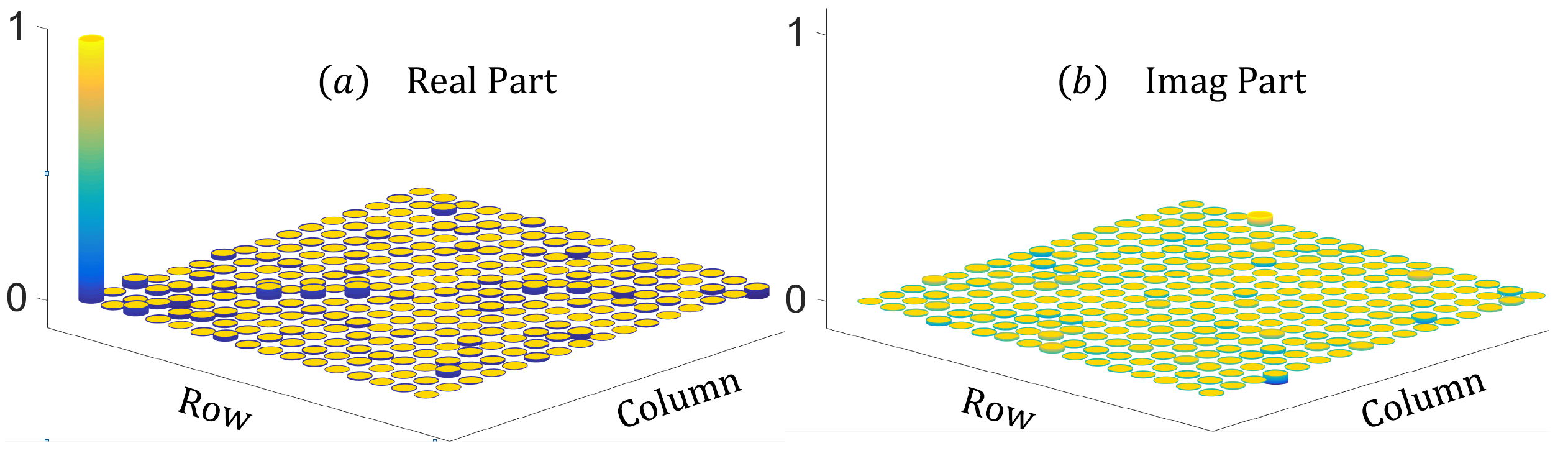}
\end{center}
\setlength{\abovecaptionskip}{-0.00cm}
\caption{\footnotesize{\textbf{Real and imaginary parts of the reconstructed PPS matrix.}  (a) and (b) respectively show the real and imaginary elements of the PPS matrix reconstructed in the experiments. The $x$ and $y$ axes represent the index number of the row and columns of the PPS matrix from 1 to 16. The $z$ axis shows the value of each element of the PPS matrix.}}\label{pps}
\end{figure*}
\\

{\it{\bfseries{Computation of the error bars}}}-- In small scale experimental setups, a good characterisation of the error sources can be useful to estimate the confidence interval of the measured expectation values, with a reduced number of experimental runs. In this respect, we follow a standard procedure that goes as follows. For each experimental realization, we numerically simulate the GRAPE pulse including a good decoherence model of our qubits. We compare the expectation values computed in this manner with the measured ones. From their discrepancy, which in average was found to be of $2.71\%$ for the three-qubit simulator and $2.35\%$ for the four-qubit one, plus the $1.30\%$ error associated to the infidelity of the initial state preparation, we estimate bounds for the experimental error of each point.  We then assume a Gaussian distribution that will yield values inside these bounds with a $95\%$ probability and we give the error bars of each point associated to the width of this Gaussian distribution. 

For experiments of bigger size, where the numerical simulation of the experiment cannot be of assistance to compute the error bars, one would increase the number of experimental runs, in order to have a statistically significant amount of data from which a reliable value of the variance can be computed.


\begin{thebibliography}{99}

\bibitem{Amico_ent} L. Amico, R. Fazio, A. Osterloh, and V. Vedral, Rev. Mod. Phys.  {\bf 80}, 517 (2008).

\bibitem{Horodecki_ent} R. Horodecki, P. Horodecki, M. Horodecki, and K. Horodecki, Rev. Mod. Phys.  {\bf 81}, 865 (2009).

\bibitem{chuang} M. A. Nielsen and I. Chuang, {\it Quantum Computation and Quantum Information} (Cambridge University Press, Cambridge, England, 2000).

\bibitem{Vidal99} G. Vidal, J.Mod.Opt. {\bf 47}, 355 (2000).

\bibitem{cramer} M. Cramer, M. B. Plenio, S. T. Flammia, R. Somma, D. Gross, S. D. Bartlett, O. Landon-Cardinal, D. Poulin, and Y.-K. Liu, Nat. Commun.  {\bf 1}, 149 (2010).

\bibitem{jullien} T. Jullien, P. Roulleau, B. Roche, A. Cavanna, Y. Jin, and D. Glattli, Nature  {\bf 514}, 603 (2014).

\bibitem{daweixin} T. Xin, D. Lu, J. Klassen, N. Yu, Z. Ji, J. Chen, X. Ma, G.-L. Long, B. Zeng, R. Laflamme, Phys. Rev. Lett. {\bf 118}, 020401 (2017).

\bibitem{park} H. S. Park, S. S. B. Lee, H. Kim, S. K. Choi, and H. S. Sim, Phys. Rev. Lett. {\bf 105}, 230404 (2010).

\bibitem{Bovino} F. A. Bovino, G. Castagnoli, A. Ekert, P. Horodecki, C. M. Alves, and A. V. Sergienko, Phys. Rev. Lett. {\bf 95}, 240407 (2005).

\bibitem{Huber} M. Huber, F. Mintert, A. Gabriel, and B. C. Hiesmayr, Phys. Rev. Lett. {\bf 104}, 210501 (2010).

\bibitem{xindawei} D. Lu, T. Xin, N. Yu, Z. Ji, J. Chen, G.-L. Long, J. Baugh, X. Peng, B. Zeng, R. Laflamme, Phys. Rev. Lett. {\bf 116}, 230501 (2016).

\bibitem{Candia} R. Di Candia, B. Mejia, H. Castillo, J. S. Pedernales, J. Casanova, and E. Solano, Phys.Rev. Lett. {\bf 111}, 240502 (2013).

\bibitem{Loredo} J. C. Loredo, M. P. Almeida, R. Di Candia, J. S. Pedernales, J. Casanova, E. Solano, and A. G. White, Phys. Rev. Lett. {\bf 116},  070503 (2016).

\bibitem{Siewert05} A. Osterloh and J. Siewert, Phys. Rev. A {\bf 72}, 012337 (2005).

\bibitem{Jones} J. A. Jones, V. Vedral, A. Ekert, G. Castagnoli, Nature  {\bf 403}, 6772 (2000).

\bibitem{Hollenberg13} M. W. Doherty, N. B. Manson, P. Delaney, F. Jelezko, J. Wrachtrup and L. C. L. Hollenberg, Phys. Reports {\bf 528}, 1 (2013).

\bibitem{Wineland03} D. Leibfried, R. Blatt, C. Monroe, and D. Wineland, Rev. Mod. Phys. {\bf 75}, 281 (2003).

\bibitem{Pedernales14} J. S. Pedernales, R. Di Candia, P. Schindler, T. Monz, M. Hennrich, J. Casanova, and E. Solano, Phys. Rev. A {\bf 90}, 012327 (2014).

\bibitem{Plenio17} M. Abdi, M.-J. Hwang, M. Aghtar, M. B. Plenio, arXiv:1704.00638.

\bibitem{sma} See the Supplemental Material for additional details.

\bibitem{cory1} D. G. Cory, A. F. Fahmy, and T. F. Havel, PNAS  {\bf 94}, 1634 (1997).

\bibitem{cory2} D. G. Cory, M. D. Price, and T. F. Havel, Physica D  {\bf 120}, 82 (1998).

\bibitem{dawei2} D. Lu, N. Xu, R. Xu, H. Chen, J. Gong, X. Peng, and J. Du, Phys. Rev. Lett. {\bf 107}, 020501 (2011).

\bibitem{Leskowitz} G. M. Leskowitz and L. J. Mueller, Phys. Rev. A  {\bf 69}, 052302 (2004).

\bibitem{JSlee} J.-S. Lee, Phys. Lett. A  {\bf 305}, 349 (2002).

\bibitem{xin15} T. Xin, H. Li, B.-X. Wang, and G.-L. Long, Phys. Rev. A {\bf 92}, 022126 (2015).

\bibitem{refc} M. R. Bendall and R. E. Gordon, J. Magn. Reson.  {\bf 53}, 365 (1983).

\bibitem{Khaneja} N. Khaneja, T. Reiss, C. Kehlet, T. Schulte-Herbruggen, and S. J. Glaser, J. Magn. Reson. {\bf 172}, 296 (2005).

\bibitem{Ryan} C. A. Ryan, C. Negrevergne, M. Laforest, E. Knill, and R. Laflamme, Phys. Rev. A {\bf 78}, 012328 (2008).

\end{thebibliography}
\end{document}